\documentclass[11pt,a4paper]{article}
\usepackage[hyperref]{emnlp-ijcnlp-2019}
\usepackage{times}
\usepackage{latexsym}
\usepackage{url}
\usepackage{graphicx}
\usepackage{amsmath}
\usepackage{booktabs}
\usepackage{algorithm}
\usepackage{algorithmic}
\usepackage{amsfonts}
\usepackage{subfigure}
\usepackage{color}

\aclfinalcopy 

\title{A Deep Neural Information Fusion Architecture for Textual \\Network Embeddings}

\author{
	Zenan Xu$^{1,2}$, \quad Qinliang Su$^{1,2}$\thanks{~~Corresponding author.}~, \quad Xiaojun Quan$^1$, \quad Weijia Zhang$^3$\\
	$^1$School of Data and Computer Science, Sun Yat-sen University \\
	$^2$Guangdong Key Laboratory of Big Data Analysis and Processing, Guangzhou, China\\
	$^3$School of Electronics and Information Technology, Sun Yat-sen University\\
	\texttt{xuzn@mail2.sysu.edu.cn, suqliang@mail.sysu.edu.cn}\\
	\texttt{quanxj3@mail.sysu.edu.cn, zhangwj39@mail2.sysu.edu.cn} \\
}

\date{}

\begin{document}
\maketitle
\begin{abstract}
Textual network embeddings aim to learn a low-dimensional representation for every node in the network so that both the structural and textual information from the networks can be well preserved in the representations. Traditionally, the structural and textual embeddings were learned by models that rarely take the mutual influences between them into account. In this paper, a deep neural architecture is proposed to effectively fuse the two kinds of informations into one representation. The novelties of the proposed architecture are manifested in the aspects of a newly defined objective function, the complementary information fusion method for structural and textual features, and the mutual gate mechanism for textual feature extraction. Experimental results show that the proposed model outperforms the comparing methods on all three datasets.
\end{abstract}

\section{Introduction}

Networks provide an effective way to organize heterogeneous relevant data, which can often be leveraged to facilitate downstream applications. For example, the huge amount of textual and relationship data in social networks contains abundant information on people's preferences, and thus can be used for  personalized advertising and recommendation. To this end, traditionally a matrix representing the network structure is often built first, then subsequent tasks proceed. However, matrix methods are computationally expensive and cannot be applied to large-scale networks.

Network embedding (NE) maps every node of a network into a low-dimensional vector, while seeking to retain the original network information. Subsequent tasks ({\em e.g.} similar vertices search, linkage prediction) can proceed by leveraging these low-dimensional features. To obtain network embeddings, \cite{perozzi2014deepwalk} proposed to first generate sequences of nodes by randomly walking along connected nodes. Word embedding methods are then employed to produce the embeddings for nodes by noting the analogies between node sequences and sentences in natural languages. Second-order proximity information is further taken into account in \cite{tang2015line}. To gather the network connection information more efficiently, the random walking strategy in \cite{perozzi2014deepwalk} is modified to favor the important nodes in \cite{grover2016node2vec}. However, all these methods only took the network structure into account, ignoring the huge amount of textual data. In Twitter social network, for example, tweets posted by a user contain valuable information on the user's preferences, political standpoints, and so on \cite{Bandyopadhyay2018SaC2VecIN}. 

To include textual information into the embeddings, \cite{tu2017cane} proposed to first learn  embeddings for the textual data and network structure respectively, and then concatenate them to obtain the embeddings of nodes. The textual and structural embeddings are learned with an objective that encourages embeddings of neighboring nodes to be as similar as possible. Attention mechanism is further employed to highlight the important textual information by taking the impacts of texts from neighboring nodes into account. Later, \cite{shen2018improved} proposed to use the fine-grained word alignment mechanism to replace the attention mechanism in \cite{tu2017cane} in order to absorb the impacts from neighboring texts more effectively. However, both methods require the textual and structural embeddings from neighboring nodes to be as close as possible even if the nodes share little common contents. This could be problematic since a social network user may be connected to users who post totally different viewpoints because of different political standpoints. If two nodes are similar, it is the node embeddings, rather than the individual textual or structural embeddings, that should be close. Forcing representations of dissimilar data to be close is prone to yield bad representations. Moreover, since the structural and textual embeddings contain some common information, if they are concatenated directly, as done in \cite{tu2017cane,shen2018improved}, the information contained in the two parts is entangled in some very complicated way, increasing the difficulties of learning representative network embeddings.

In this paper, we propose a novel deep neural {\bf I}nformation {\bf F}usion {\bf A}rchitecture for textual {\bf N}etwork {\bf E}mbedding (NEIFA) to tackle the issues mentioned above. Instead of forcing the separate embeddings of structures and texts from neighboring nodes to be close, we define the learning objective based on the node embeddings directly. For the problem of information entanglement, inspired by the gating mechanism of long short-term memory (LSTM) \cite{hochreiter1997long}, we extract the complementary informations from texts and structures and then use them to constitute the node embeddings. A mutual gate is further designed to highlight the node's textual information that is consistent with neighbors' textual contents, while diminishing those that contracdict to each other. In this way, the model provides a mechanism to only allow the information that is consistent among neighboring nodes to flow into the node embeddings. The proposed network embedding method is evaluated on the tasks of link prediction and vertex classification, using three real-world datasets from different domains. It is shown that the proposed method outperforms state-of-the-art network embedding methods on the task of link prediction by a substantial margin, demonstrating that the obtained embeddings well retain the information in original networks. Similar phenomenons can also be observed in the vertex classification task. These results suggest the effectiveness of the proposed neural information fusion architecture for textual network embeddings.

\section{Related Work}

\begin{figure*}[ht]
	\centering
	\includegraphics[scale=0.65]{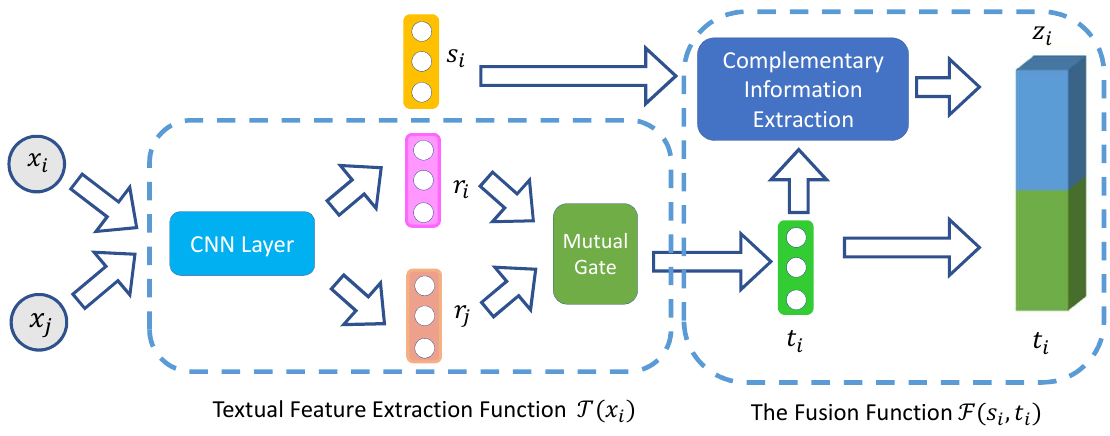}
	\caption{The overall framework of NEIFA.}
	\label{fig:model}
\end{figure*}

\noindent {\bf Text Embedding} \quad There has been various methods to embed textual information into vector representations for NLP tasks. The classical method for embedding textual information could be one-hot vector, term frequency inverse document frequency (TF-IDF), etc. Due to the high-dimension and sparsity problems in here, \cite{mikolov2013efficient} proposed a novel neural network based skip-gram model to learn distributed word embeddings via word co-occurrences in a local window of textual content. To exploit the internal structure of text, convolutional neural networks (CNNs) \cite{blunsom2014convolutional,kim2014convolutional} is applied to obtain latent features of local textual content. Then, by following a pooling layer, fixed-length representations are generated. To have the embeddings better reflect the correlations among texts, soft attention mechanisms \cite{bahdanau2014neural,vaswani2017attention} is proposed to calculate the relative importances of words in a sentence by evaluating their relevances to the content of comparing sentences. Alternatively, gating mechanism is applied to strengthen the relevant textual information, while weakening the irrelevant one by controlling the information-flow path of a network in \cite{dauphin2017language,zhou2017selective}.

\noindent {\bf Network Embedding} \quad Network embedding methods can be categorized into two classes: (1) methods that solely utilize structure information; and (2) methods that consider both structure and textual content associated with vertices. For the first type of methods, DeepWalk \cite{perozzi2014deepwalk} was the first to introduce neural network technique into the network embedding field. In DeepWalk, node sequences are generated via randomly walking on the network, dense latent representations are by feeding those node sequences into the skip-gram model. LINE \cite{tang2015line} exploited the first-order and second-order proximity information of vertices in network by optimizing the joint and condition probability of edges. Further, Node2Vec \cite{grover2016node2vec} proposed a biased random walk to search a network and generate node sequences based on the depth-first search and width-first search. However, those methods only embed the structure information into vector representations, while ignoring the informative textual contents associated with vertices. To address this issue, some recent work seeks to the joint impact of structure and textual contents to obtain better representations. TADW \cite{yang2015network} proved that DeepWalk is equivalent to the matrix factorization and the textual information can be incorporated by simply adding the textual feature into the matrix factorization. CENE \cite{sun2016general} extends the original networks by transforming the textual content into another kinds of vertices, and the vertices are embedded into low-dimensional representations on the extended network. CANE \cite{tu2017cane} proposed to learn separate embeddings for the textual and structural information, and obtain the network embeddings by simply concatenating them, in which a mutual attention mechanism is used to model the semantic relationship between textual contents. WANE \cite{shen2018improved} modified the semantic extraction strategy in CANE by introducing a fine-grained word alignment technique to learn word-level semantic information more effectively. However, {most of recent} methods force the textual and structural embeddings of two neighboring nodes close to each other irrespective of their underlying contents.

\section{The Proposed Method}
A textual network is defined as $G$ = \{$V$, $E$, $T$\}, where $V$, $E$ and $T$ denote the vertices in the graph, edges between vertices and textual content associated with vertices, respectively. Each edge $e_{i,j} \in E$ suggests there is a relationship between vertex $v_i$ and $v_j$.


\subsection{Training Objective}
Suppose the structural and textual features of node $i$ are given and are denoted as $s_i$ and $t_i$, respectively. Existing methods are built on the objectives that encourage the structural and textual features of neighboring nodes to be as similar as possible. As discussed in the previous sections, this may make the node embeddings deviate from the true information in the nodes. In this paper, we define the objective based on the node embeddings  directly, that is,

\begin{equation}
	\label{lossRepresentation}
	\mathcal{L} = \sum_{\{i, j\} \in E} \log p(h_i|h_j),
\end{equation}
where $h_i$ is the network embedding of node $i$, and is constructed from $s_i$ and $t_i$ by
\begin{equation}
	h_i = {\mathcal{F}}(s_i, t_i);
\end{equation}
${\mathcal{F}}(\cdot, \cdot)$ is the fusion function that maps the structural and textual features into the network embeddings; and $p(h_i|h_j)$ denotes the conditional probability of network embedding $h_i$ given the network embedding $h_j$. Following LINE \cite{tang2015line}, the conditional probability in \eqref{lossRepresentation} is defined as:
\begin{equation}
	\label{conditionalProbabilityH}
	p(h_i|h_j) = \frac{\exp (h_i \cdot h_j)}{\sum_{z \in V} \exp (h_i \cdot h_z)}.
\end{equation}

Note that the structural feature $s_i$ is randomly initialized and will be learned along with the other model parameters. For the textual feature $t_i$, it is obtained via a trainable feature extraction function from the given texts, i.e.,
\begin{equation}
	t_i = {\mathcal{T}}(x_i),
\end{equation}
where $x_i$ represents the texts associated with node $i$. From the definition of objective function (1), it can be seen that it is the network embeddings of nodes, rather than the individual structural or textual embeddings, that are encouraged to be close for neighboring nodes. 

Details on how to realize the fusion function $\mathcal{F}(\cdot, \cdot)$ and feature extraction function ${\mathcal{T}}(\cdot)$ are deferred to Section \ref{Sec:fuse} and Section \ref{Sec:text}, respectively. The overall framework of our proposed NEIFA is shown in Fig.\ref{fig:model}.

\subsection{Fusion of Structural and Textual Features ${\mathcal{F}}(s_i, t_i)$}
\label{Sec:fuse}
\begin{figure}[!tp]
	\centering
	\includegraphics[scale=0.1]{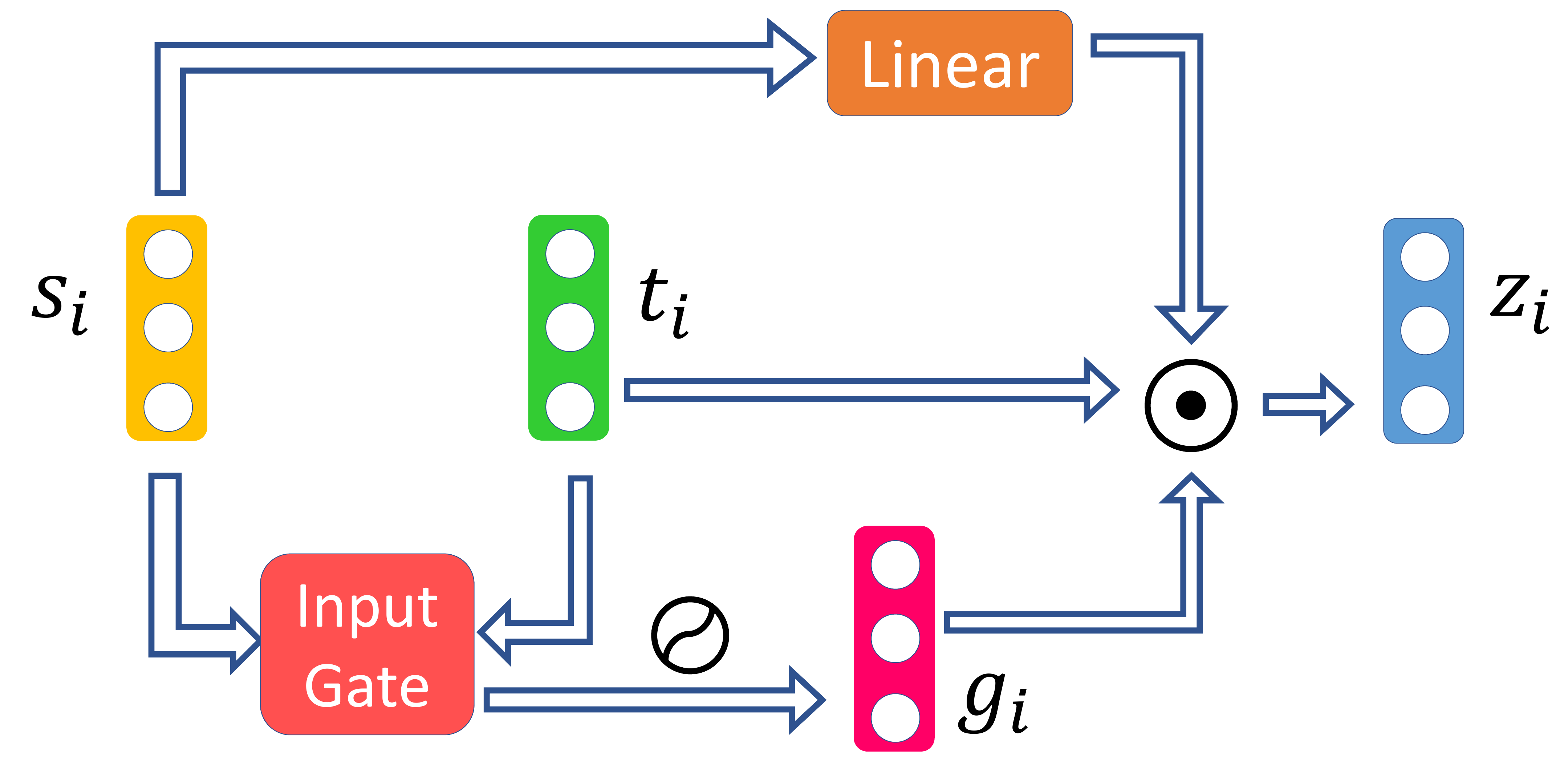}
	\caption{The Fusion Module.}
	\label{fig:complementary}
\end{figure}

In this section, we will present how to fuse the structural and textual features to yield the embeddings for nodes. The fusion module ${\mathcal{F}}(s_i, t_i)$ is illustrated in Fig.\ref{fig:complementary}. The simplest way to obtain network embeddings is to concatenate them directly, {\em i.e.} $h_i=[s_i;t_i]$. However, it is known that the structural and textual features are not fully exclusive, and often contain some common information. Thus, if the network embeddings are generated by simply concatenating the two features, different parts of the embeddings become entangled to each other in some unknown but complex way. This may make the process of optimizing the objective function more difficult and hinder the model to learn representative embeddings for the nodes. In this paper, we instead distill the information that is complementary to the textual feature $t_i$ from $s_i$ first, and then concatenate the two complementary information to constitute the embeddings of nodes.

To distill the complementary information from the structural feature $s_i$, inspired by LSTM, an input gate is designed to eliminate the information in $s_i$ that has already appeared in $t_i$. Specifically, the gate is designed as
\begin{equation}
	g_i = 1- \sigma(({\mathbf{P}}s_i + b_g) \odot t_i),
\end{equation}
where $\sigma(\cdot)$ is the sigmoid function; $\odot$ denotes the element-wise multiplication; ${\mathbf{P}}$ and $b_g$ are used to align the structural feature $s_i$ to the space of textual features $t_i$. From the definition of $g_i$, it can be seen that  if the values on some specific dimension of ${\mathbf{P}}s_i + b_g$ and $t_i$ are both large, which indicates the same information appears in both $s_i$ and $t_i$, the gate $g_i$ will be closed. So, if $({\mathbf{P}}s_i + b_g)\odot t_i$ is multiplied to the gate $g_i$, only the information that is not contained in both $s_i$ and $t_i$ is allowed to pass through. Thus, $(({\mathbf{P}}s_i + b_g) \odot t_i)\odot g_i$ can be understood as the information in $s_i$ that is complementary to $t_i$. In practice, we untie the values of ${\mathbf{P}}$ and $b_g$, and use a new trainable matrix ${\mathbf{Q}}$ and bias $b_c$ instead. The complementary information is eventually computed as
\begin{equation}
	z_i = (({\mathbf{Q}}s_i + b_c) \odot t_i)\odot g_i. 
\end{equation}
Then, we concatenate complementary information $z_i$ to the textual features $t_i$ to produce the final network embedding

\begin{equation}
	h_i = [z_i;t_i].
\end{equation}
In this way, given the structural and textual features $s_i$ and $t_i$, we successfully extract the complementary information from $s_i$ and generate the final network embedding $h_i$.

\subsection{Textual Feature Extraction ${\mathcal{T}(x_i)}$}
\label{Sec:text}

When extracting textual features for the embeddings of nodes, the impacts from neighboring nodes should also be taken into account, {\em i.e.} highlighting the consistent information, while dampening the inconsistent ones. To this end, we first repersent words with their corresponding embeddings, and then apply a one-layer CNN followed by an average pooling operator to extract the raw features for texts \cite{tu2017cane, shen2018baseline}. Given the raw textual features $r_i$ and $r_j$ of two neighboring nodes $i$ and $j$, we diminish the information that are not consistent in the two raw features. Specifically, the final textual features are computed for nodes $i$ and $j$ as

\begin{equation}
	\label{iGatej}
	t_{i} = r_i \odot \sigma(r_j),
\end{equation}
\begin{equation}
	\label{jGatei}
	t_{j} = r_j \odot \sigma(r_i),
\end{equation}
where $\sigma(\cdot)$ serves as the role of gating. Since the raw textual feature $r_i$ often exhibits specific meanings on different dimensions, the expressions \eqref{iGatej} and \eqref{jGatei} can be understood as a way to control which information is allowed to flow into the embeddings. Only the information that is consitent among neighboring nodes can appear in the textual feature $t_i$, which is then fused into the network embeddings. There are a variety of other nonlinear functions that can serve as the role of gating, but in this work, the simplest but effective sigmoid function is employed.

\subsection{Training Details}

Maximizing the objective function in \eqref{lossRepresentation} requires to compute the expensive softmax function repeatedly, in which the summation over all nodes of the networks is needed for iteration. To address this issue, for each edge $e_{i,j} \in E$, we introduce negative sampling \cite{mikolov2013distributed} to simplify the optimization process. Therefore, the conditional distribution $p(h_i|h_j)$ into the following form:

\begin{equation}
	\log \sigma (h_i\cdot h_j) + \sum_{k=1}^{K}E_{h_k\sim P(h)} [\log \sigma(-h_k\cdot h_i)]
\end{equation}
where $K$ is the number of negative samples, and $P(v) \propto d_v^{3/4}$ is the distribution of vertices with $d_v$ representing the out-degree of vertex $v$. Adam \cite{kingma2014adam} is employed to optimize the entire model based on randomly mini-batch of edges in each step.

\section{Experiments}
To evaluate the quality of the network embeddings generated by the proposed method, we apply them in two tasks: link prediction and vertex classification. Link prediction aims to predict whether there exists a link between two randomly chosen nodes based on the similarities of embeddings of the two nodes. Vertex classification, on the other hand, tries to classify the nodes into different categories based on the embeddings, provided that there exists some supervised information. Both tasks can achieve good performances only when the embeddings retain important information of the nodes, including both the structural and textual information. In the following, we will first introduce the datasets and baselines used in this paper, then describe the evaluation metric and experimental setups, and lastly report the performance of the proposed model on the tasks of link prediction and vertex classification, respectively.

\subsection{Datasets and Baselines}
\begin{table}[!htp]
	\centering
	\begin{tabular}{l|rrr}
		\toprule
		Datasets  & {Zhihu} & {Cora} & {HepTh} \\
		\midrule
		Vertices & 10000 & 2277 & 1038 \\
		Edges & 43894 & 5214 & 1990  \\
		\#(Edges) & 2191 & 93 & 24 \\
		*(Text) & 190 & 90 & 54\\
		Labels & - & 7 & - \\
		\bottomrule
	\end{tabular}
	\caption{Statistics of datasets, where \#(Edges) denotes the max number of the connective relationship of a node, and *(Text) denotes the average lengths of the text.}
	\label{table:datasets}
\end{table}

Experiments are conducted on three real-world datasets: Zhihu \cite{sun2016general}, Cora \cite{mccallum2000automating} and HepTh \cite{leskovec2005graphs}. Below shows the detailed descriptions of the three datasets, with their statistics summaries given in Table \ref{table:datasets}. {The preprocessing procedure of the above datasets is the same as that in \cite{tu2017cane} \footnote[1]{https://github.com/thunlp/CANE}.}

\begin{itemize}
	\item {\emph{Zhihu}} \cite{sun2016general} is a Q\&A based community social network. In our experiment, 10000 active users and the descriptions of their interested topics are collected as the vertices and texts of the social network to be studied. There are total 43894 edges which indicate the relationship between active users.
	\item {\emph{Cora}} \cite{mccallum2000automating} is a citation network that consists of 2277 machine learning papers with text contents divided into 7 categories. The citation relations among the papers are reflected in the 5214 edges.
	\item {\emph{HepTh}} \cite{leskovec2005graphs} (High Energy Physics Theory) is a citation network from the e-print arXiv. In our experiment, 1038 papers with abstract information are collected, among which 1990 edges are observed.
\end{itemize}

To evaluate the effectiveness of our proposed model, several strong baseline methods are compared with, which are divided into two categories as follows:
\begin{itemize}
	\item {\emph{Structure-only:}} DeepWalk \cite{perozzi2014deepwalk}, LINE \cite{tang2015line}, Node2vec \cite{grover2016node2vec}.
	\item {\emph{Structure and Text:}} TADW \cite{yang2015network}, CENE \cite{sun2016general}, CANE \cite{tu2017cane}, WANE \cite{shen2018improved}.
\end{itemize}

\subsection{Evaluation Metrics and Experimental Setups}
\begin{table*}[t]
	\centering
	\begin{tabular}{lrrrrrrrrr}
		\toprule
		\% Training Edges & 15\% & 25\% & 35\% &45 \% & 55\% & 65\% & 75\% & 85\% & 95\% \\
		\midrule
		DeepWalk$^\dag$ & 56.6 & 58.1 & 60.1 & 60.0 & 61.8 & 61.9 & 63.3 & 63.7 & 67.8 \\
		LINE$^\dag$ & 52.3 & 55.9 & 59.9 & 60.9 & 64.3 & 66.0 & 67.7 & 69.3 & 71.1 \\
		node2vec$^\dag$ & 54.2 & 57.1 & 57.3 & 58.3 & 58.7 & 62.5 & 66.2 & 67.6 & 68.5 \\
		\midrule
		TADW$^\dag$ & 52.3 & 54.2 & 55.6 & 57.3 & 60.8 & 62.4 & 65.2 & 63.8 & 69.0 \\
		CENE$^\dag$ & 56.2 & 57.4 & 60.3 & 63.0 & 66.3 & 66.0 & 70.2 & 69.8 & 73.8 \\
		CANE$^\dag$ & 56.8 & 59.3 & 62.9 & 64.5 & 68.9 & 70.4 & 71.4 & 73.6 & 75.4 \\
		WANE$^\ddag$ & 58.7 & 63.5 & 68.3 & 71.9 & 74.9 & 77.0 & 79.7 & 80.0 & 82.6 \\
		\midrule
		NEIFA & {\bf 68.9} & {\bf 73.7} & {\bf 78.3} & {\bf 81.0} & {\bf 84.5} & {\bf 87.3} & {\bf 88.2} & {\bf 89.6} & {\bf 90.1} \\
		
		\bottomrule
	\end{tabular}
	\caption{AUC scores for link prediction on Zhihu. Note that $\dag$ and $\ddag$ indicate the results are taken from \cite{tu2017cane} and \cite{shen2018improved}, respectively.}
	\label{table:Zhihu}
\end{table*}

\begin{table*}[t]
	\centering
	\begin{tabular}{lrrrrrrrrr}
		\toprule
		\% Training Edges & 15\% & 25\% & 35\% &45 \% & 55\% & 65\% & 75\% & 85\% & 95\% \\
		\midrule
		DeepWalk$^\dag$ & 56.0 & 63.0 & 70.2 & 75.5 & 80.1 & 85.2 & 85.3 & 87.8 & 90.3 \\
		LINE$^\dag$ & 55.0 & 58.6 & 66.4 & 73.0 & 77.6 & 82.8 & 85.6 & 88.4 & 89.3 \\
		node2vec$^\dag$ & 55.9 & 62.4 & 66.1 & 75.0 & 78.7 & 81.6 & 85.9 & 87.3 & 88.2 \\
		\midrule
		TADW$^\dag$ & 86.6 & 88.2 & 90.2 & 90.8 & 90.0 & 93.0 & 91.0 & 93.4 & 92.7 \\
		CENE$^\dag$ & 72.1 & 86.5 & 84.6 & 88.1 & 89.4 & 89.2 & 93.9 & 95.0 & 95.9 \\
		CANE$^\dag$ & 86.8 & 91.5 & 92.2 & 93.9 & 94.6 & 94.9 & 95.6 & 96.6 & 97.7 \\
		WANE$^\ddag$ & {\bf 91.7} & {\bf 93.3} & 94.1 & 95.7 & 96.2 & 96.9 & 97.5 & 98.2 & 99.1 \\
		\midrule
		NEIFA & 89.0 & 92.2 & {\bf 95.3} & {\bf 96.5} & {\bf 97.1} & {\bf 97.4} & {\bf 97.6} & {\bf 98.5} & {\bf 99.2} \\
		\bottomrule
	\end{tabular}
	\caption{AUC scores for link prediction on Cora. Note that $\dag$ and $\ddag$ indicate the results are taken from \cite{tu2017cane} and \cite{shen2018improved}, respectively.}
	\label{table:Cora}
\end{table*}

In link prediction, the performance criteria of area under the curve (AUC) \cite{hanley1982meaning} is used, which represents the probability that vertices in a random unobserved link are more similar than those in a random non-existent link.

For the vertex classification task, a logistic regression model is first trained to classify the embeddings into different categories based on the provided labels of nodes. Then, the trained model is used to classify the network embeddings in test set, and the classification accuracy is used as the performance criteria of this task.

To have a fair comparison with competitive methods, the dimension of network embeddings is set to 200 for all considered methods. The number of negative samples is set to 1 and the mini-batch size is set to 64 to speed up the training processes. Adam \cite{kingma2014adam} is employed to train our model with a learning rate of $1 \times 10^{-3}$. 

\subsection{Link Prediction}

\begin{table*}[!htp]
	\centering
	\begin{tabular}{lrrrrrrrrr}
		\toprule
		\% Training Edges & 15\% & 25\% & 35\% &45 \% & 55\% & 65\% & 75\% & 85\% & 95\% \\
		\midrule
		DeepWalk$^\dag$ & 55.2 & 66.0 & 70.0 & 75.7 & 81.3 & 83.3 & 87.6 & 88.9 & 88.0 \\
		LINE$^\dag$ & 53.7 & 60.4 & 66.5 & 73.9 & 78.5 & 83.8 & 87.5 & 87.7 & 87.6 \\
		node2vec$^\dag$ & 57.1 & 63.6 & 69.9 & 76.2 & 84.3 & 87.3 & 88.4 & 89.2 & 89.2 \\
		\midrule
		TADW$^\dag$ & 87.0 & 89.5 & 91.8 & 90.8 & 91.1 & 92.6 & 93.5 & 91.9 & 91.7 \\
		CENE$^\dag$ & 86.2 & 84.6 & 89.8 & 91.2 & 92.3 & 91.8 & 93.2 & 92.9 & 93.2 \\
		CANE$^\dag$ & 90.0 & 91.2 & 92.0 & 93.0 & 94.2 & 94.6 & 95.4 & 95.7 & 96.3 \\
		WANE$^\ddag$ & {\bf 92.3} & 94.1 & 95.7 & 96.7 & {\bf 97.5} & 97.5 & 97.7 & 98.2 & 98.7 \\
		\midrule
		NEIFA & 91.7 & {\bf 94.2} & {\bf 95.9} & {\bf 96.8} & 97.4 & {\bf 97.6} & {\bf 98.0} & {\bf 98.6} & {\bf 99.1} \\
		\bottomrule
	\end{tabular}
	\caption{AUC scores for link prediction on HepTh. Note that $\dag$ and $\ddag$ indicate the results are taken from \cite{tu2017cane} and \cite{shen2018improved}, respectively.}
	\label{table:HepTh}
\end{table*}

\begin{figure*}[htbp]
	\centering
	\subfigure[Zhihu]{
		\begin{minipage}[t]{0.32\linewidth}
			\centering
			\includegraphics[width=1.95in]{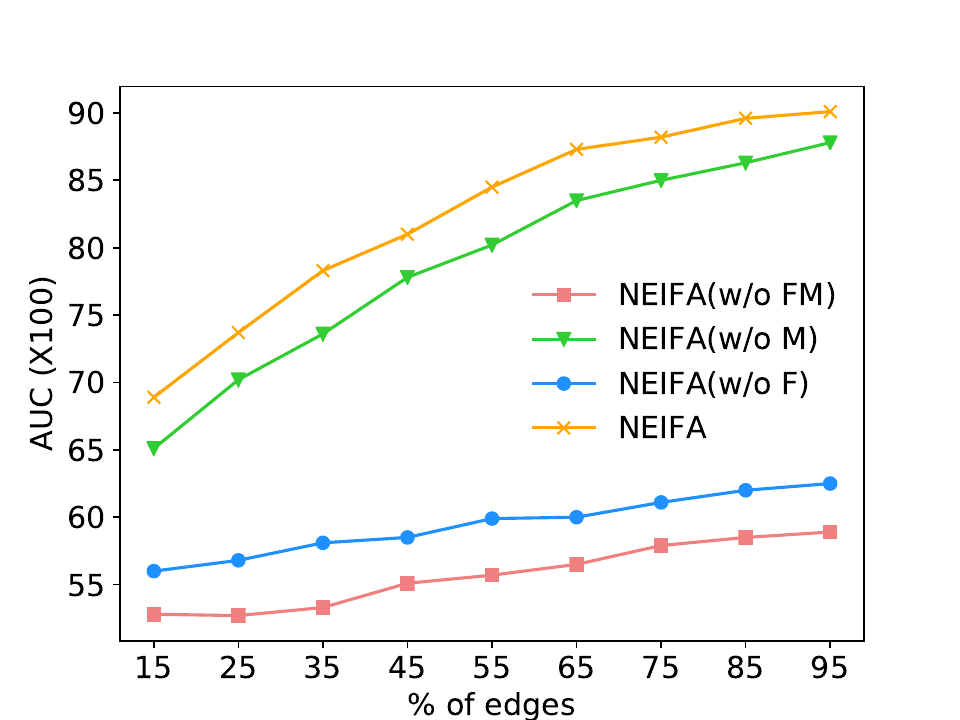}
		\end{minipage}
	}%
	\subfigure[Cora]{
		\begin{minipage}[t]{0.32\linewidth}
			\centering
			\includegraphics[width=2.0in]{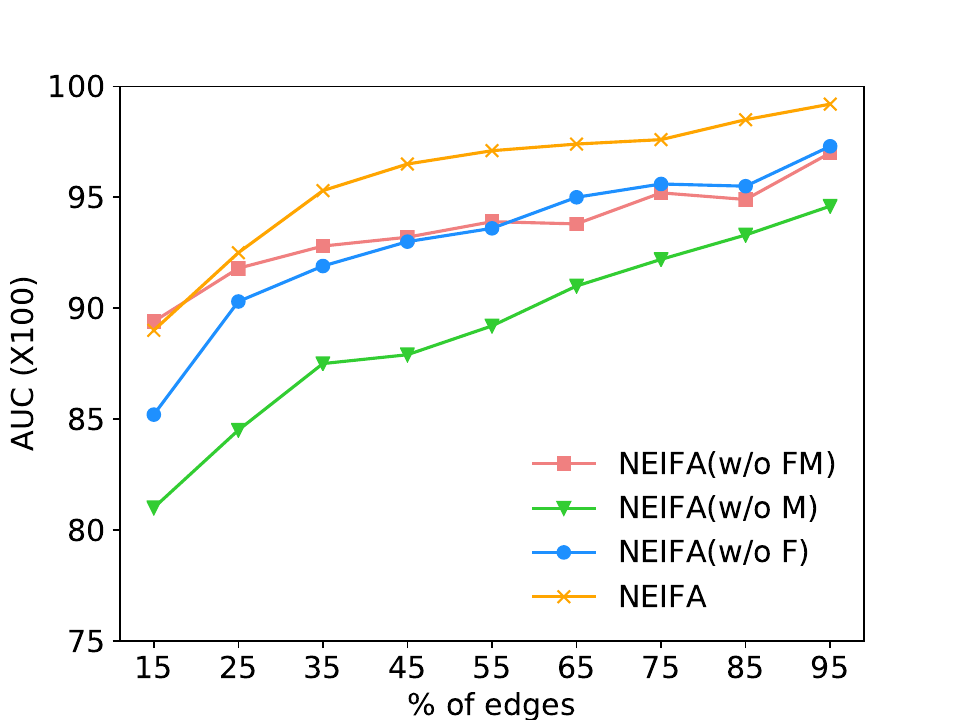}
		\end{minipage}%
	}%
	\centering
	\subfigure[HepTh]{
		\begin{minipage}[t]{0.32\linewidth}
			\centering
			\includegraphics[width=1.98in]{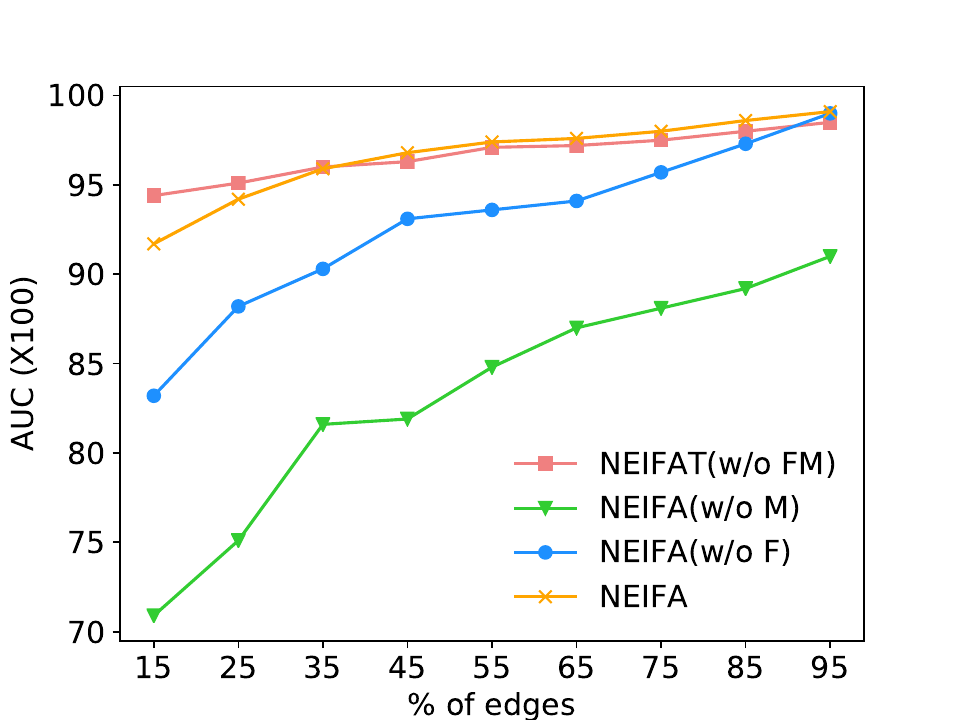}
		\end{minipage}%
	}%
	\caption{Ablation study of NEIFA model on different datasets on link prediction task.}
	\label{fig:ablation}
\end{figure*}

We randomly extract a portion of edges from the whole edges to constitute the training datasets, and use the rest as the test datasets. The AUC scores of different models under proportions ranging from 15\% to 95\% on Zhihu, Cora and HepTh datasets are shown in Table \ref{table:Zhihu}, Table \ref{table:Cora} and Table \ref{table:HepTh}, respectively, with the best performance highlighted in bold.

As can be seen from Table \ref{table:Zhihu}, our proposed method outperforms all other baselines in Zhihu dataset substantially, with approximately a 10 percent improvement over the current state-of-the-art WANE model. This may be partially attributed to the complicated Zhihu dataset, in which both the structures and texts contain important informations. If the two individual features are concatenated directly, there may be sever information overlapping problem, limiting the models to learning good embeddings. The proposed complementary information fusion method alleviate the issue by disentangling the structural and textual features. In adition, the proposed mutual gate mechanism that removes inconsistent textual information from a node's textual feature also contribute to the performance gains. 
On the other hand, the substantial gain may also be partially attributed to the objective function that is directly defined on the network embeddings. That is because the inconsistencies of the structural or textual information among neighboring nodes are more likely to happen in complex networks. 

For the other two datasets, as shown in Table \ref{table:Cora} and Table \ref{table:HepTh}, our proposed method outperforms baseline methods overall. The results strongly demonstrate that the network embeddings generated by the proposed model are easier to preserve the original information in the nodes. It can be also seen that  the performance gains observed in the Cora and HepTh datasets are not as substantial as that in Zhihu dataset. The relatively small improvement may be attributed to the fact that the number of edges and neighbors in Cora and HepTh datasets are much smaller that Zhihu datasets. We speculate that the information in structures of the two datasets is far less than that in texts, implying that the overlapping issue is not as sever as that in Zhihu. Hence, direct concatenation will not induce significant performance loss.

\noindent {\bf Ablation Study} \quad To demonstrate the effectiveness of proposed fusion method and mutual gate mechanism, three variants of the proposed model are evaluated: (1)NEIFA(w/o FM): NEIFA without both fusion process and mutual gated mechanism where the raw textual features $r$ are directly regarded as network embeddings. (2) NEIFA(w/o F): NEIFA without fusion process where the textual features $t$ are directly regarded as network embeddings. (3) NEIFA(w/o M): NEIFA without mutual gated mechanism where the network embeddings are obtained by fusing the structural features and raw textual features. The three variants are compared with original NEIFA model on the three datasets above. The results are showed in Fig.\ref{fig:ablation}. It can be seen that for networks with very sparse structure, such as Hepth, the method that simply uses the raw textual features as their network embeddings can achieve pretty good performance. In the simple datasets, the proposed model even exhibits worse performance in the case of small proportion of training edges. As the datasets become larger and more complex network structure is included, the performance of only using the textual embeddings decreases rapidly. The reason may be that as the networks grow, the differences of structural or textual data among neighboring nodes become more apparent, and the advantages of the mutual gate mechanism and information fusion method will show up.

\subsection{Vertex Classification}
To demonstrate the superiority of proposed method, the vertex classification experiment is also considered on the Cora dataset. This experiment is established on the basis that if the original network contains different types of nodes, good embeddings mean that they can be classified into specific classes by a simple classifier easily. For the proposed method, the embedding of a node varies as it interacts with different nodes. To have the embedding fixed, we follow procedures in \cite{tu2017cane} to yield a node's embedding by averaing the embeddings that are obtained when the node interacts with different neighbors.

\begin{figure}[t!]
	\centering
	\includegraphics[scale=0.45]{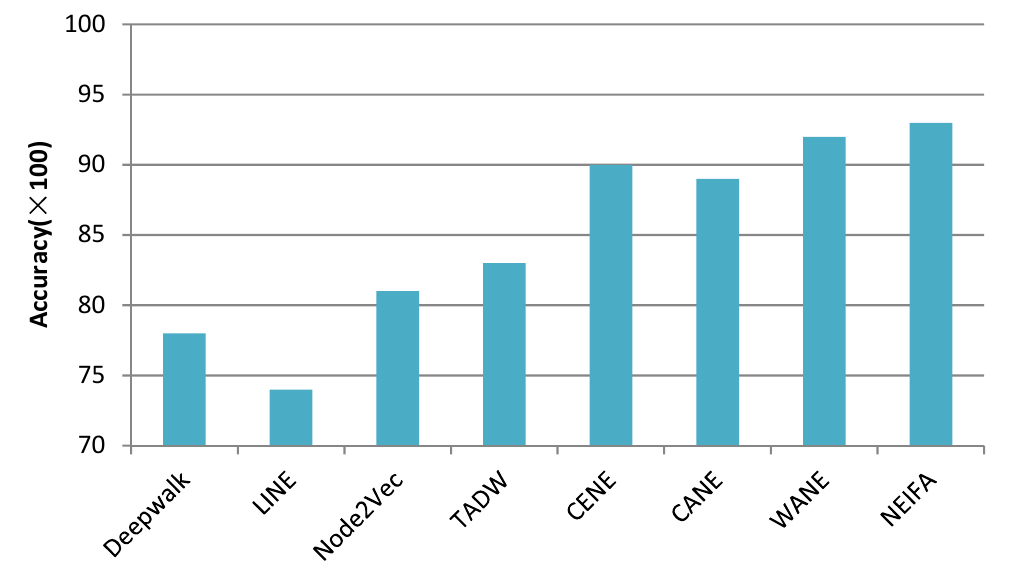}
	\caption{Vertex classification result on Cora dataset}
	\label{fig:bar}
\end{figure}

To this end, we randomly split the node embeddings of all nodes with a proportion of 50\%-50\% into a training and testing set, respectively. A logistic regression classifier regularized by $L_2$ distance \cite{fan2008liblinear} is then trained on the node embeddings from training set. The classification performance is tested on the hold-out testing set. The above procedures are repeated 10 times and their average value is reported as the final performance. It can be seen from Fig.\ref{fig:bar} that methods considering both structural and textual information show better classification accuracies than methods leveraging only structural information, demonstrating the importance of incorporating textual information into the embeddings. Moreover, NEIFA outperforms all methods considered, which further proves the superiority of our proposed model.


\begin{figure}[t!]
	\centering
	\includegraphics[scale=0.33]{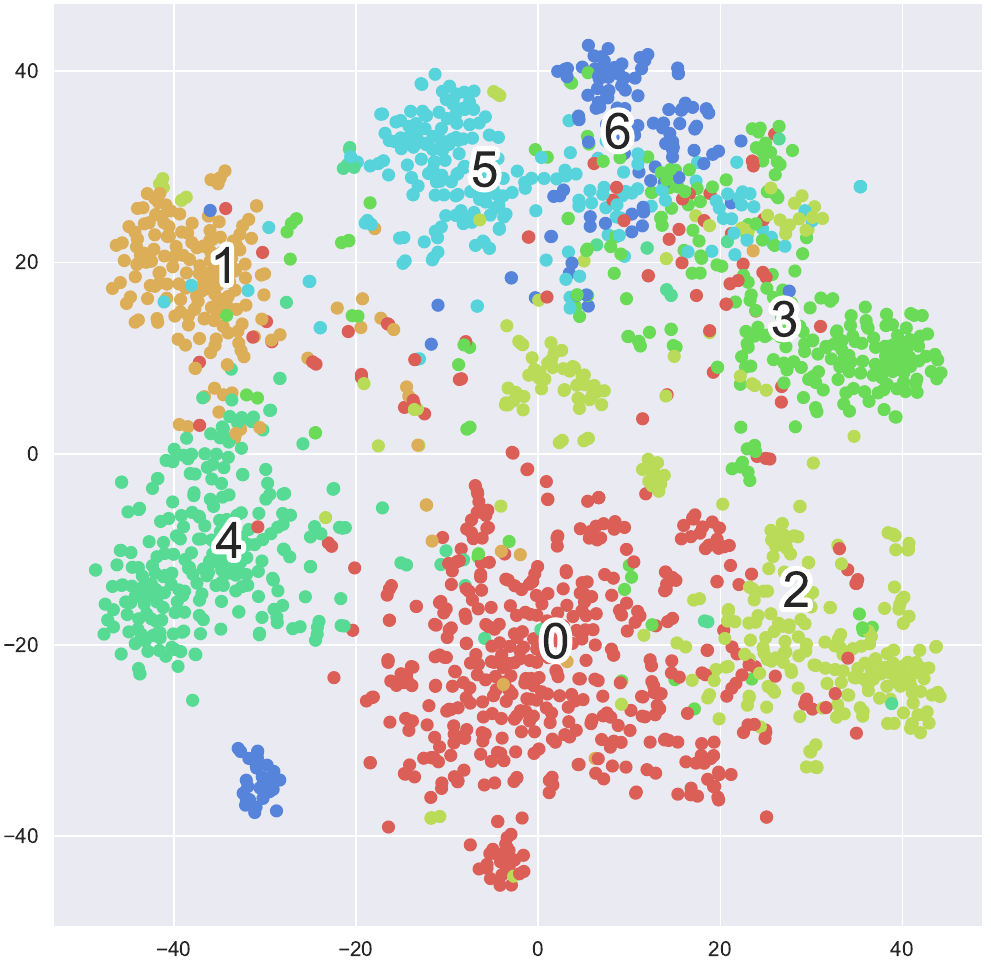}
	\caption{$t-$SNE visualization of our learned network embeddings on Cora dataset.}
	\label{fig:tsen}
\end{figure}

To intuitively understand the embeddings produced by the proposed model, we employ $t$-SNE to map our learned embeddings to a 2-D space. The result is shown in Fig.\ref{fig:tsen}, where different colors indicate that the nodes belong to different categories. Note that, although the mapping in t-SNE is trained without using any category labels, the latent label information is still partially extracted out. As shown in Fig.\ref{fig:tsen}, the points with the same color are closer to each other, while the ones with different colors are far apart.

\section{Conclusions}
In this paper, a novel deep neural architecture is proposed to effectively fuse the structural and textual informations in networks. Unlike existing embeddings methods which encourage both textual and structural embeddings of two neighboring nodes close to each other, we define the training objective based on the node embeddings directly. To address the information duplication problem in the structural and textual features, a complementary information fusing method is further developed to fuse the two features. Besides, a mutual gate is designed to highlight the textual information in a node that is consistent with the textual contents of neighboring nodes, while diminishing those that are conflicting to each other. Exhaustive experimental results on several tasks manifest the advantages of our proposed model.

\section*{Acknowledgments}
We thank Wei Liu and Wenxuan Li for their helps. This work is supported by the National Natural Science Foundation of China (NSFC) under Grant No. 61806223, U1711262, U1501252, U1611264 and U1711261, and National Key R\&D Program of China (2018YFB1004404).

\newpage
\bibliography{emnlp-ijcnlp-2019}
\bibliographystyle{acl_natbib}

\appendix

\end{document}